# Plasma-induced magnetic phase in 3D Mn$^{II}$-Nb$^{IV}$ octacyanidometalate with magnetic sponge behavior

Dominik Czernia[a]*, Piotr Konieczny[a], Marcin Perzanowski[a], Beata Nowicka[c], Dawid Pinkowicz[c]

[a]Institute of Nuclear Physics PAN, ul. Radzikowskiego 152, 31-342 Cracow, Poland
[b]Faculty of Chemistry, Jagiellonian University, ul. Gronostajowa 2, 30-387 Cracow, Poland
*Corresponding author: dominik.czernia@ifj.edu.pl

**Abstract**

A new magnetic phase with $T_C$ = 72 K was obtained by exposing the three-dimensional $\{[Mn^{II}(H_2O)_2]_2[Nb^{IV}(CN)_8]\cdot 4H_2O\}_n$ coordination ferrimagnet ($T_C$ = 49 K) to air, oxygen, nitrogen, and argon-based plasma. The X-ray powder diffraction pattern revealed that the unit cell shrank after plasma treatment, leading to a 20% enhancement of the superexchange couplings, as estimated from the mean-field approximation (MFA) model. Although no stable dehydrated form was found in the thermogravimetric analysis, the observed changes are attributed to the removal of crystallization water molecules. The plasma-induced magnetic phase could not be obtained by exposing the studied material to 0% relative humidity during dynamic vapor sorption. Instead, the material underwent a major structural reorganization after dehydration, necessitating an extended MFA model to reproduce the magnetic susceptibility. These findings demonstrate that plasma-induced changes can create unique magnetic phases in molecule-based systems that are otherwise unobtainable.

1. Introduction

Molecular magnetic materials show great potential for modern applications that focus on molecule-based properties, such as spintronics [1] and qubit designs [2], magnetic refrigeration [3,4], or bistable switches and active sensors [5]. This wide range of possibilities emerges from a variety of available systems that are sensitive to external stimuli such as temperature, pressure [6], light irradiation [7], or the incorporation of guest molecules [8–10]. The primary approach of the molecular magnetism community is to engineer materials at the molecular level by selecting appropriate synthetic building blocks and procedures [11]. Additionally, alternative post-synthetic methods have been successfully applied, including the deposition of molecular magnets on thin films [12] or ion beam irradiation [12].

Plasma treatment is a cost-effective and fast technique used in surface etching [13,14], functionalization [15], polymerization [16], and polymer surface activation and grafting [17]. Plasma can introduce specific chemical and structural changes that can lead to the alteration of magnetic properties, such as changes in the net magnetic moment [18–21], magnetic hysteresis loops [22–25], and blocking temperatures in nanoparticles [26,27]. Even though molecular magnets and polymers share a common chemical composition and bonding, there have been few, if any, reports on the combination of plasma treatment and molecular magnetism.

To fill this gap, we conducted a study to investigate the impact of plasma irradiation on the magnetic properties of a three-dimensional $\{[Mn^{II}(H_2O)_2]_2[Nb^{IV}(CN)_8]\cdot 4H_2O\}_n$ octacyanidometalate soft ferrimagnet with a Curie temperature of $T_C = 49$ K. This coordination polymer is a member of a well-studied isostructural family of compounds [28–30], known for exhibiting magnetocaloric effect [31–34], magnetic percolation [35], or magnetic sponge behavior [36,37]. Our study considered plasma generated from atmospheric air, nitrogen, oxygen, and argon with varying irradiation time and plasma chamber power. As a result, we observed an increase in $T_C$ to 72 K. Our findings, based on X-ray powder diffraction and mean-field approximation model, indicated that the original core crystallographic structure is conserved after plasma treatment with only minor contraction of the unit cell, which was enough to enhance the superexchange coupling constants by 20%. We ruled out any potential effect of generated heat and ultraviolet (UV) light on the observed $T_C$ shift, as well as the possible dehydration of the system. We concluded that the obtained magnetic phase was solely attributed to the studied material interaction with plasma.

**2. Materials and methods**

2.1 Sample preparation

The $\{[Mn^{II}(H_2O)_2]_2[Nb^{IV}(CN)_8]\cdot 4H_2O\}_n$ molecular system, denoted as **NbMn$_2$**, was synthesized using $MnCl_2\cdot 4H_2O$ and $K_4[Nb(CN)_8]\cdot 2H_2O$ as described in [35], forming a dark red powder. Manual mortar was used to decrease the powder grain size, which was then spread on a 1×1 cm$^2$ Scotch tape strip, increasing the surface-to-volume ratio. The mass of the thin, transparent orange layer on the tape strip was approximately 0.2 mg. Immediately after exposure to the plasma, heat, or UV light, the samples were protected from air by placing another layer of the tape strip on top and encapsulating them inside a gelatin capsule sealed with high vacuum Apiezon M grease.

The reference sample underwent a similar procedure, including grounding, spreading it on a tape strip, and putting it in the vacuum for 30 minutes inside the plasma chamber. The sample used in the dynamic vapor sorption (DVS) experiment was prepared differently, without scotch tape. Instead, 4.6 mg of powder was used during the DVS, which was protected in a sealed film afterward.

2.2 Thermogravimetric analysis (TGA)

A thermogravimetric analysis (TGA) was performed on a 1.702 mg sample in an argon atmosphere from 20 °C to 250 °C with a 2 °C/min heating rate. The selected high temperature limit is sufficient for observing the $CN^-$ decomposition in other octacyanidometalates.

2.3 Plasma, heat, and ultraviolet (UV) treatment

The samples were exposed to plasma generated by the Harrick Plasma Cleaner with approximately 1 mbar gas pressure and varying parameters: plasma cleaner power (low — 7 W, medium — 11 W, high — 18 W), gas type used (atmospheric air, oxygen, nitrogen, argon), and treatment time (2, 10, 15 minutes). After plasma treatment using 11 W/18 W power for at least 10 minutes, the color of the samples changed from orange to dark brown. The temperature during treatment was monitored with an infrared thermometer and did not exceed 75 °C.

Three additional **NbMn$_2$** samples were investigated to ensure the observed effects were exclusively due to plasma treatment. Two samples were exposed to 90 °C and 150 °C for 30 minutes in an oven under 60 mbar pressure. Another sample was exposed to $\lambda = 395$ nm light generated by a 35 W UV LED flashlight for 30 minutes in atmospheric air.

2.4 X-ray powder diffraction (XRPD)

PANalytical X'Pert Pro instrument equipped with a copper X-ray tube source (Cu K$_{\alpha 1}$, 1.541 Å) was used to obtain X-ray powder diffraction (XRPD) patterns in the $\theta$-$2\theta$ geometry, operating at 40 kV and 30 mA. The data was recorded at room temperature in the $2\theta$ range of 5-55° and corrected by the background correction and the removal of the anode's characteristic X-ray K$_{\alpha 2}$.

2.5 Magnetometry

The Quantum Design SQUID MPMS-XL magnetometer was used to measure the magnetic properties. The isothermal magnetization $M(\mu_0 H)$ at $T = 2.0$ K was obtained in the applied field of $\mu_0 H$ = [-7 T, 7 T]. The static magnetic susceptibility $\chi$ was measured in $\mu_0 H_{dc}$ = 500 G from 300 K to 2 K and additionally in two modes: field cooling (FC) and zero-field cooling (ZFC) in $\mu_0 H_{dc}$ = 100 G in the $T$ = 2-100 K temperature range. The temperature-independent diamagnetic contribution $\chi_0$ from the studied compound and sample holder was subtracted from the measured signals by fitting Curie-Weiss law with $\chi_0$ to the $\chi(T)$ dependence obtained in $\mu_0 H_{dc}$ = 500 G. The temperature dependence of in-phase $\chi'$ ac magnetic susceptibility was obtained using a 3 G amplitude and 120 Hz frequency of alternating magnetic field at 2-80 K.

## 3. Results

3.1 X-ray powder diffraction (XRPD)

The **NbMn₂** molecular magnet crystallizes in the tetragonal space group I4/m, forming a three-dimensional network that features Nb$^{IV}$ and two Mn$^{II}$ metal centers in a unit cell, connected through Nb$^{IV}$–CN–Mn$^{II}$ linkages (Fig. 1a). The system contains four water of crystallization molecules. Two other water molecules are coordinated to the Mn$^{II}$ ion. As depicted in Fig. 1b, the Nb$^{IV}$ metal center has an approximately square antiprismatic coordination, and the Mn$^{II}$ metal center has a slightly distorted octahedron coordination. Two crystallographically independent Nb$^{IV}$-Mn$^{II}$ connections have 5.539 Å and 5.482 Å straight-line distances between the metal centers. The bond angles Nb$^{IV}$–C–N are almost linear, ranging between 176.3° and 177.1°, while the Mn$^{II}$–N–C bond angles are between 154.2° and 166.4°. For more details regarding the **NbMn₂** structure, refer to [28] report.

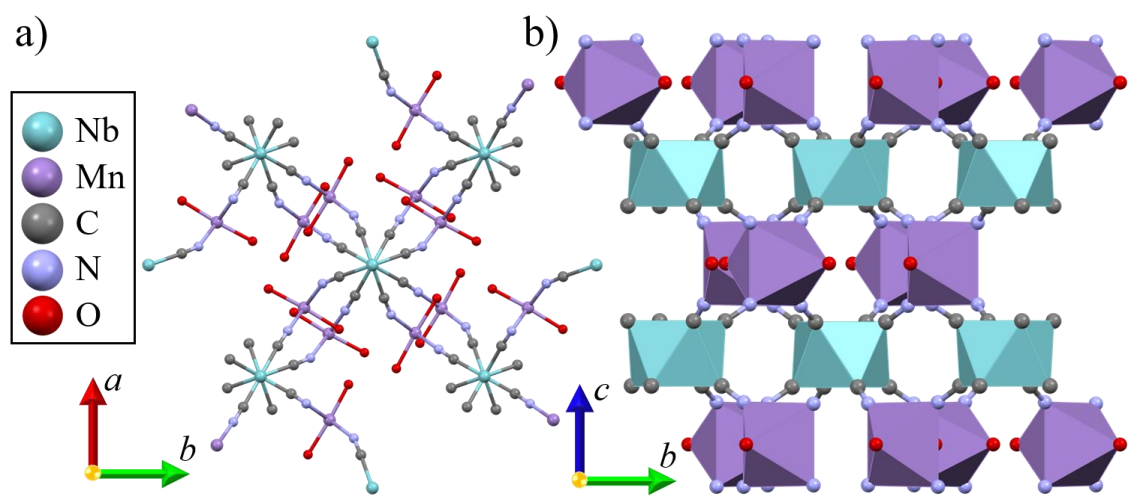

Figure 1. The crystal structure of **NbMn₂**: (a) ball and stick representation along the *c* crystallographic axis; (b) polyhedral representation along *a* crystallographic axis. Water of crystallization and hydrogen atoms are omitted for clarity.

The comparison between XRPD patterns obtained for the reference sample, sample irradiated with air-based high-power plasma for 15 minutes, samples heated at 90 °C and 150 °C, and a scotch tape strip is shown in Fig. 2. The reference and plasma-treated samples have the same peaks. Still, for the latter, the peaks are shifted by 0.1-0.2° towards higher 2θ, which may reflect the change in interplane distances up to 0.01 Å according to Bragg's law. The Williamson-Hall plots analysis of the selected peaks showed a slight reduction in crystallite size from $D = 98(16)$ nm to $D = 57(13)$ nm after plasma irradiation. The lattice microstrains ε were within the uncertainty level: $\varepsilon = 2.22(35)\times10^{-3}$ for the reference and $\varepsilon = 2.65(77)\times10^{-3}$ for plasma-irradiated samples.

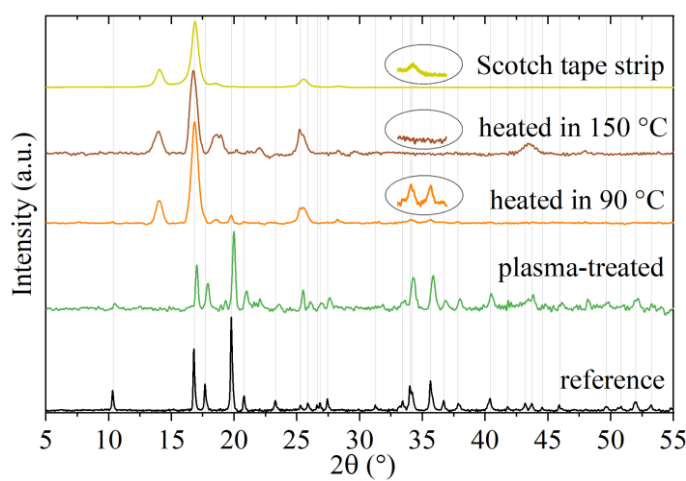

Figure 2. XRPD patterns measured in the 2θ range of 5-55° for the following samples (from the bottom): reference, plasma-treated, heated at 90 °C and 150 °C, Scotch tape strip. The circles enlarge the 33-37° area for the corresponding curves. Gray lines indicate peak maxima found for the reference sample.

The XRPD patterns of the samples heated in an oven showed a substantial contribution from the tape strip, indicating a partial loss of the sample, most likely due to the evaporation of the glue on the tape. For the sample heated at 90 °C, peaks corresponding to the reference XRPD pattern were identified at approximately 10.3°, 19.8°, 20.8°, 25.2°, 34°, and 35.7°. These peaks were not shifted as **NbMn$_2$** after plasma treatment. The XRPD pattern for the sample heated at 150 °C showed no clear correspondence with the reference's pattern. Additionally, new peaks appeared at around 19°, 20.2°, 22°, 28.3°, 43.5°, and 47.9°, revealing the decomposition of the original **NbMn$_2$** compound.

## 3.2 Thermogravimetric analysis (TGA)

The variation in mass $m$ measured during TGA for **NbMn$_2$** is presented in Fig. 3. An initial upward drift in $m$ by 3.7% between 20 °C and 50 °C results from small sample mass and possible confinement of argon in the intergranular spaces. Further decrease in $m$ by 9.4% from 50 °C to 250 °C shows no apparent plateau in this temperature range, suggesting that no stable dehydrated form of **NbMn$_2$** can be obtained by heating.

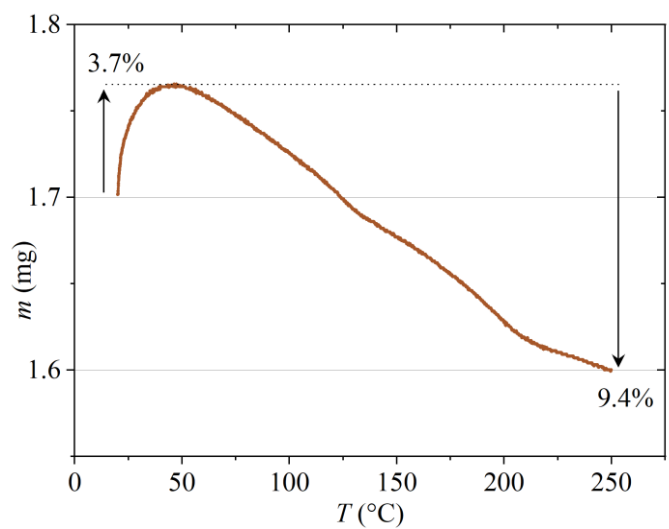

Figure 3. Thermogravimetric analysis for the **NbMn$_2$** sample with the initial mass $m$ = 1.702 mg performed in an argon atmosphere in the 20-250 °C range (heated 2 °C per minute).

## 3.3 Dynamic vapor sorption (DVS)

The DVS of **NbMn$_2$** showed up to 22% net mass change with narrow hysteresis related to the possible sorption and desorption (Fig. 4) of eight water molecules. These include four crystallization and four coordinated water molecules. The process has two vague steps with no evident structural transition. The expected net mass change for incorporating four water molecules is approximately 15%, which is observed at 80% relative humidity. The further increase in net mass change may be due to surface sorption effects or the incorporation of two additional water molecules. The latter should exhibit a 24% net mass change, which is close to the experimental value of 22%.

If six water molecules are removed, i.e., all water of crystallization and half of the coordinated water, each Mn ion would presumably be coordinated by only one water molecule instead of two. This type of dehydration would cause lattice reorganization.

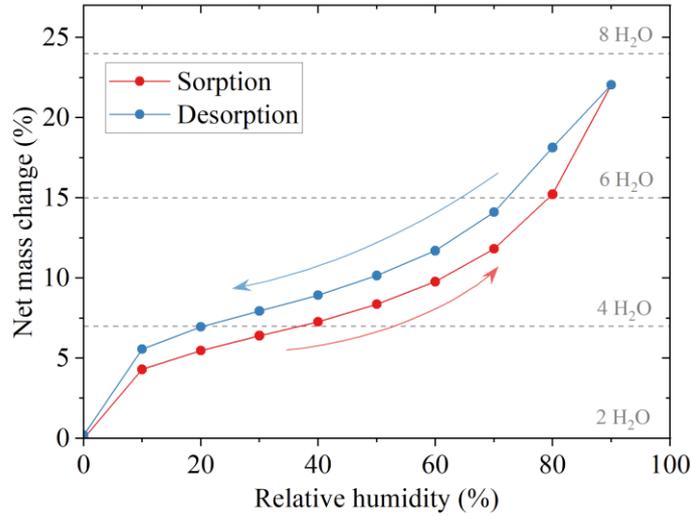

Figure 4. The **NbMn₂** sample net mass change in dynamic vapor sorption experiment during sorption (red points) and desorption (blue points). Grey dashed lines indicate the expected number of water molecules in the formula unit for specific net mass change, assuming two water molecules are in the crystal lattice at 0% relative humidity. Solid lines are a guide to the eye.

3.3 Magnetic properties

The **NbMn₂** compound has been previously investigated and described in another article [28]. It is a soft ferrimagnet without a magnetic hysteresis loop, exhibiting long-range magnetic order below the Curie temperature of $T_C$ = 49 K. The antiferromagnetic interaction between one $Nb^{IV}$ ($S$ = 1/2) and two $Mn^{IV}$ ($S$ = 5/2) magnetic centers per formula unit is reflected by saturation magnetization of $M_S$ = 9 $\mu_B \cdot mol^{-1}$, which is already achieved in approximately $\mu_0 H$ = 1 T (Fig. 5). This value is expected, assuming a $g$ factor of 2.0 for both ions.

The effects of **NbMn₂** sample grounding were investigated before the plasma irradiation experiment. Figure 5. displays the isothermal magnetization curves obtained at $T$ = 2.0 K for three samples: as-synthesized, grounded, grounded and protected with an additional layer of a tape strip. The $M(H)$ curves were overlapping for all the studied samples. On the other hand, the magnetic susceptibility temperature dependence shape (inset in Fig. 5) obtained in the field cooling (FC) and zero-field cooling (ZFC) modes in $\mu_0 H$ = 100 G differs between the as-synthesized and grounded samples. In addition, the grounded but unprotected sample shows the presence of two magnetic phases, one of which is characterized by a shift in the Curie temperature, determined from the FC/ZFC curves' first derivative minima, from $T_C$ = 49 K to $T_C$ = 64 K signalizing the modification of magnetic interactions within the studied system. This process is time-dependent and does not occur if the sample is protected from air.

The observed shape alteration of low-field FC/ZFC curves is attributed to the reduced crystallite size due to the grounding process. This effect is negligible when investigating plasma impact on the magnetic properties of **NbMn₂**. Therefore, the grounded and protected sample will be referred to as the reference later in the text.

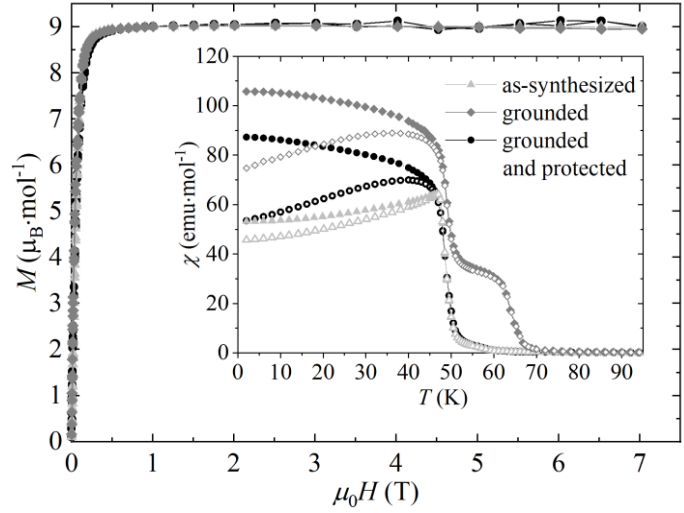

Figure 5. The isothermal magnetization measured at $T = 2.0$ K in the applied magnetic fields $\mu_0H = 0$-7 T of the as-synthesized (light gray triangles), grounded (gray diamonds), and grounded and protected (black circles) **NbMn₂** samples. The inset shows the field cooling (full points) and zero-field cooling (empty points) magnetic susceptibilities in the function of temperature ($T = 2$-95 K) measured in $\mu_0H = 100$ G for the corresponding samples. Solid lines are a guide to the eye.

The influence of plasma irradiation on the real part $\chi'$ of the ac magnetic susceptibility of **NbMn₂** is shown in Fig. 6. When the sample was treated with air-based high-power plasma for 15 minutes (Fig. 6a), a new magnetic phase was created, with $T_C = 72$ K obtained from the $\chi'(T)$ first derivative minimum. For simplicity, the plasma-treated sample used later in the text will refer to this particular sample, unless stated otherwise, to represent the complete post-plasma magnetic phase modification. Reducing the treatment time to 10 minutes showed the presence of two magnetic phases with $T_C = 48$ K and $T_C = 69$ K, corresponding to the as-synthesized and plasma-induced **NbMn₂** magnetic systems, respectively. The sample irradiated with plasma for 2 minutes showed no substantial changes in the $\chi'(T)$ dependence. Therefore, the modification process of the **NbMn₂** magnetic system induced by plasma is irradiation time-dependent, with a specific threshold time required to observe any changes.

Similarly, comparable effects were observed when pure nitrogen, oxygen, or argon were used to generate plasma at high power for 10 minutes (Fig. 6b). In all cases, two magnetic phases with $T_C = 48$ K and $T_C = 69$ K were found, but oxygen and nitrogen-based plasmas were more effective than air/argon-based plasmas when comparing the $\chi'$ values increase at both $T_C$.

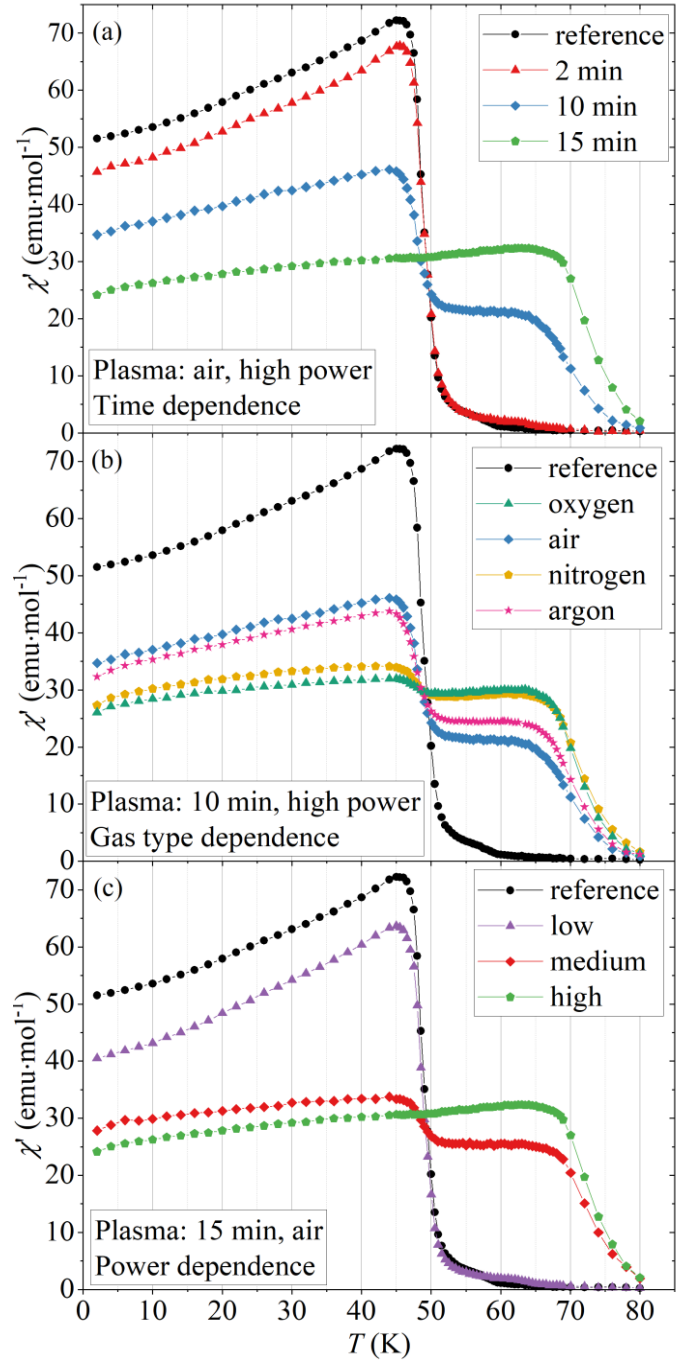

Figure 6. The temperature dependence of the real part of the ac magnetic susceptibility measured with $\mu_0 H_{ac}$ = 3 G amplitude and 120 Hz frequency of magnetic field for the reference (black circles) and plasma-treated **NbMn$_2$** samples in varying conditions. Top: air-based high-power plasma used for 2 min (red triangles), 10 min (blue diamonds), 15 min (green pentagons); middle: 10 min high-power plasma created from oxygen (turquoise triangles), air (green diamonds), nitrogen (yellow pentagons), argon (pink stars); bottom: 15 min air-based plasma created with low — 7 W (purple triangles), medium — 11 W (red diamonds), high — 18 W (green pentagons) plasma generator power. Solid lines are guide to the eye.

The effects of different power levels of a plasma cleaner were studied using air-based plasma for 15 minutes (Fig. 6c). Lower power resulted in similar effects to reducing the irradiation time (Fig. 6a). The complete and partial magnetic phase changes were observed at high and medium power levels, respectively. At a low power level, only one $T_C$ = 49 K was found, as in the reference sample. An additional measurement conducted in low-power air-based plasma for 30 minutes produced the same results. Therefore, a sufficiently high power level of the plasma generator is necessary to induce changes in the **NbMn₂** magnetic system.

Figure 7. presents isothermal magnetization curves of **NbMn₂** modified using air-based high-power plasma for 2, 10, and 15 minutes, measured at $T$ = 2.0 K in $\mu_0 H$ = [0, 7 T]. The shapes of the obtained curves were similar to the reference's, reaching an expected value of $M_S$ = 9 $\mu_B \cdot$mol$^{-1}$ in high magnetic fields. This confirmed that the core magnetic properties of **NbMn₂**, namely one Nb$^{IV}$ and two Mn$^{II}$ metal centers per formula unit, coupled antiferromagnetically, were preserved. However, a relative decrease in magnetization values was observed in lower magnetic fields, notably for the sample irradiated for 15 minutes, indicating a slight modification in the magnetization process after plasma treatment. Similar conclusions were drawn for **NbMn₂** magnetic behavior, investigated in terms of plasma gas type and plasma cleaner power dependence. In the plasma-irradiated samples, a minor widening of the narrow hysteresis loop was observed compared to the reference (inset in Fig. 7). In particular, the coercivity increased from $\mu_0 H$ = 20 Oe to $\mu_0 H$ = 60 Oe, and the remanence from $M_R$ = 0.25 $\mu_B \cdot$mol$^{-1}$ to $M_R$ = 0.6 $\mu_B \cdot$mol$^{-1}$ for the sample treated in plasma for 15 minutes.

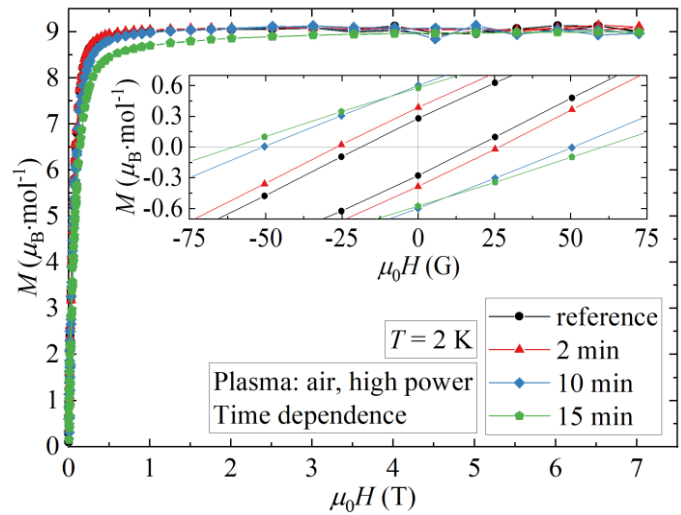

Figure 7. The isothermal magnetization measured at $T$ = 2.0 K in the applied magnetic fields $\mu_0 H$ = 0-7 T of the **NbMn₂** samples: reference (black circles), air-based high-power plasma irradiated for 2 min (red triangles), 10 min (blue diamonds), and 15 min (green pentagons). The inset shows the $M(H)$ dependence in the $\mu_0 H$ = [-75 G, 75 G] range. Solid lines are a guide to the eye.

An investigation was conducted to determine the potential effect of heat and ultraviolet (UV) light generated during the plasma treatment on the magnetic properties of **NbMn$_2$**. The magnetic susceptibility $\chi$ temperature dependence was measured from 300 K to 2 K in $\mu_0H$ = 500 G and presented as the $\chi T$ product in Fig. 8 for the reference and plasma-treated samples, and samples heated in an oven at 90 and 150 °C and exposed to UV light ($\lambda$ = 395 nm). For all samples, a peak in $\chi T(T)$ dependence was observed with the minimum of $\chi T(T)$ derivative at $T_C$ = 49 K for the reference and the samples exposed to 90 °C, and UV light, and at $T_C$ = 69 K for the samples heated in 150 °C and treated in plasma.

The $\chi T(T)$ curves for the samples exposed to heat and UV light had considerably lower values than the reference $\chi T(T)$: 3-4 times lower for the samples heated to 90 °C and irradiated with UV light and 30 times lower for the sample heated to 150 °C. In addition, there was a non-negligible $\chi T$ offset in the high-temperature range tail and an extra hump around $T$ = 20 K for the sample heated to 150 °C (inset in Fig. 8). The partial decomposition of the **NbMn$_2$** system may have caused these effects.

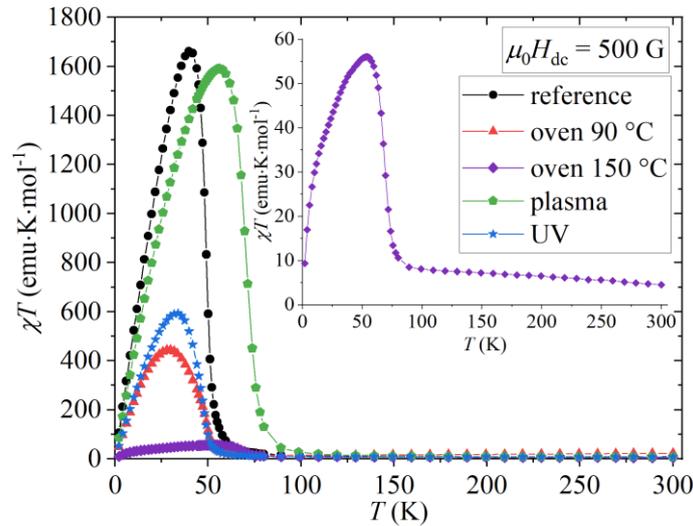

Figure 8. The temperature dependence of magnetic susceptibility and temperature product $\chi T$ measured in $\mu_0H$ = 500 G while cooling from 300 K to 2 K of the **NbMn$_2$** samples: reference (black circles), plasma irradiated (green pentagons), heated at 90 °C in an oven (salmon triangles), heated at 150 °C in an oven (purple diamonds), and exposed to ultraviolet light (UV, blue stars). The inset shows the 30-times zoomed-in chart of $\chi T$ for the sample heated at 150 °C. Solid lines are a guide to the eye.

Figure 9 displays the isothermal magnetization at $T$ = 2 K measured in varying magnetic fields $\mu_0H$ = [0, 7 T] of the samples exposed to heat and UV light. The saturation magnetization for these samples is not achieved up to $\mu_0H$ = 7 T, reaching approximately 6 $\mu_B$·mol$^{-1}$ for all three cases, i.e., 66 % of $M_S$ found for reference and plasma-treated samples. The magnetization curves for the samples heated at 90 °C and exposed to UV light are similar in shape and values, showing an abrupt increase between 0 to 0.1 T, as for the reference, and in higher magnetic fields, a slower growth, contrary to the reference. The magnetization process for the sample heated at 150 °C is different as it does not have a steep curve at low magnetic fields. This may indicate the antiparallel alignment of $Mn^{II}$ magnetic moments, paramagnetic contribution, or magnetic anisotropy enhancement. None of these cases occur in the original **NbMn₂** magnetic system. Thus, the samples heated at 90 °C and exposed to UV light most likely underwent partial, while the sample heated at 150 °C major, magnetic system decomposition or transformation.

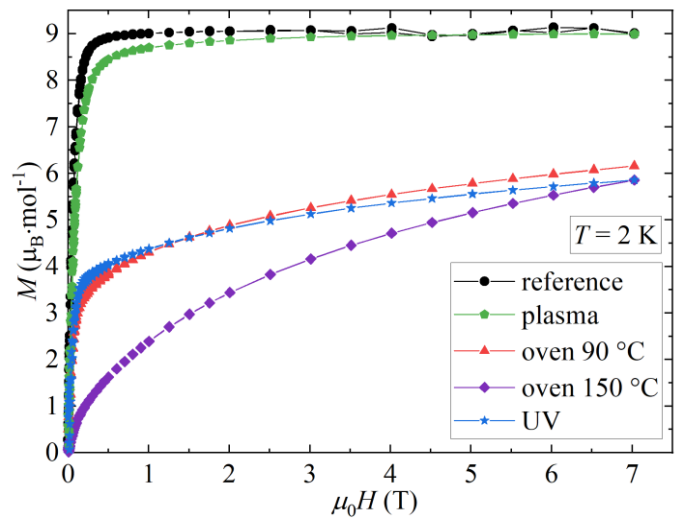

Figure 9. The isothermal magnetization measured at $T = 2.0$ K in the applied magnetic fields $\mu_0H = 0$-7 T of the **NbMn₂** samples: reference (black circles), plasma irradiated (green pentagons), heated at 90 °C in an oven (salmon triangles), heated at 150 °C in an oven (purple diamonds), and exposed to ultraviolet light, UV (blue stars). Solid lines are a guide to the eye.

Figure 10. shows the isothermal magnetization measurements at $T = 2$ K in $\mu_0H = [0, 7\text{ T}]$ for three samples: the reference, plasma-treated, and the sample exposed to 0% relative humidity during the dynamic vapor sorption (post-DVS). The latter sample did not show magnetization saturation and reached only 3.42 $\mu_B \cdot mol^{-1}$ at $\mu_0H = 7$ T. This value is approximately 40% of the expected $M_S$ for the considered system. Additionally, the coercivity increased to $\mu_0H_c = 300$ G (inset in Fig. 10), compared to $\mu_0H_c \approx 20$ G for the reference and $\mu_0H_c \approx 60$ G for the plasma-treated samples. Similarly to the sample heated at 150 °C, exposing the **NbMn₂** system to 0% relative humidity may modify the magnetization process by enhancing magnetic anisotropy or interactions leading to $Mn^{II}$ magnetic moments antiparallel alignment or causing major structural transformation.

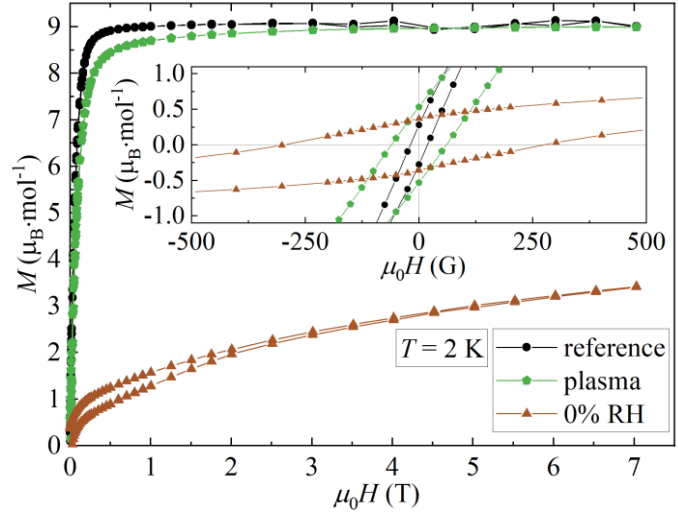

Figure 10. The isothermal magnetization measured at $T = 2.0$ K in the applied magnetic fields $\mu_0H = 0$-7 T of the **NbMn$_2$** samples: reference (black circles), plasma irradiated (green pentagons), exposed to 0% relative humidity (brown triangles). The inset shows the $M(H)$ dependence in the $\mu_0H = $ [-500 G, 500 G] range. Solid lines are a guide to the eye.

Figure 11. presents the inverse magnetic susceptibility $\chi^{-1}$ temperature dependence data measured in $\mu_0H = 500$ G for the reference, plasma-treated, and post-DVS samples. The minimum of $\chi(T)$ first derivative for the post-DVS sample indicates a shift in Curie temperature to $T_C = 40$ K, compared to $T_C = 49$ K for the reference. The corresponding Curie temperature for the plasma-treated sample is $T_C = 70$ K. The magnetic susceptibility $\chi$ for the sample exposed to 0% relative humidity has considerably lower values. This suggests a new magnetic phase with a lower Curie temperature was created, most likely due to water molecule desorption, which differs from the magnetic phase created during plasma treatment.

The evaluation of the magnetic interactions between the magnetic centers in the **NbMn$_2$** for the reference, plasma-treated, and post-DVS samples was done by employing the molecular field approximation model given by the following formula [35]:

$$\chi = \frac{\chi_{Mn} + \chi_{Nb} + 2\chi_{Mn}\chi_{Nb}\Lambda_{NbMn} - \chi_{Mn}\chi_{Nb}\Lambda_{MnMn}}{1 - \chi_{Mn}\Lambda_{MnMn} - \chi_{Mn}\chi_{Nb}\Lambda_{NbMn}^2} \quad (1),$$

where $\chi_x = N_A\mu_B^2\lambda_x g_x^2 S_x(S_x + 1)/3k_BT$ is the paramagnetic molar susceptibility of the sublattice $x =$ Nb, Mn with $N_A$ — Avogadro constant, $\mu_B$ — Bohr magneton, $\lambda_x$ — stoichiometric factors ($\lambda_{Nb} = 1$, $\lambda_{Mn} = 2$), $g_x$ — Landé factor, $S_x$ — spin quantum number ($S_{Nb} = 1/2$, $S_{Mn} = 5/2$), and $k_B$ — Boltzmann constant. $\Lambda_{xy}$ denotes the molecular field constants for the internal magnetic field emerging from sublattice $y$ acting on sublattice $x$, which read:

$$\Lambda_{NbMn} = \frac{J_{NbMn}Z_{NbMn}}{N_A\mu_B^2\lambda_{Mn}g_{Nb}g_{Mn}}, \qquad \Lambda_{MnMn} = \frac{J_{MnMn}Z_{MnMn}}{N_A\mu_B^2\lambda_{Mn}g_{Mn}^2} \qquad (2),$$

where $J_{xy}$ — superexchange coupling constant between $x$ and $y$ type ions, and $Z_{xy}$ — number of the nearest neighbor $y$ type ions around the $x$ type ion ($Z_{NbMn}$ = 8). $J_{MnMn}Z_{MnMn}$ product is considered as one parameter due to difficulties in specifying the exact $Z_{MnMn}$ value.

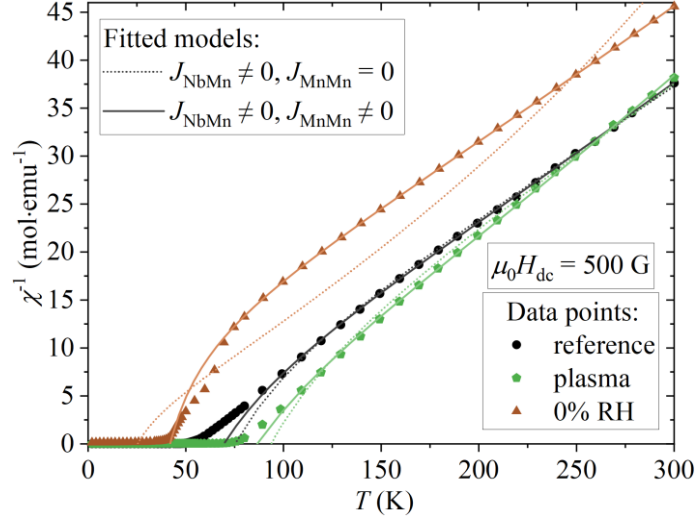

Figure 11. The inverse magnetic susceptibility temperature dependence ($T$ = 2-300 K) measured in $\mu_0H$ = 500 G for the **NbMn$_2$** reference (black circles), plasma-treated sample (green pentagons), and sample exposed to 0% relative humidity (brown triangles). Solid and dotted lines represent the best fit to the molecular field approximation (MFA) models, assuming no interactions ($J_{MnMn}$ = 0) and existing interactions ($J_{MnMn} \neq 0$), respectively, between Mn$^{II}$ magnetic centers.

The fitting procedure was split into two stages, as reported in 2018 [35]. In the first stage, the interaction between Mn$^{II}$ magnetic centers was omitted, i.e., $Z_{MnMn}J_{MnMn}$ was fixed to zero. In the second stage, the $Z_{MnMn}J_{MnMn}$ parameter was included. The fitting function used was ($\chi$ + $\chi_0$)$^{-1}$, where $\chi_0$ represents the temperature-independent magnetic susceptibility. The Landé factors of $g_{Nb} = g_{Mn} = 2.0$ were used for both stages, which is supported by the observed value of saturation magnetization of $M_S$ = 9 $\mu_B$·mol$^{-1}$ for the reference (Fig. 5). The fitting procedure was performed in the temperature range of 100-300 K using *Orthogonal Distance Regression* iteration algorithm within *Origin Pro* software. The results are presented in Fig. 11 (solid lines) and Table 1.

Table 1. Results of the mean-field approximation (MFA) model fitting to the $(\chi + \chi_0)^{-1}$ function obtained from the magnetic susceptibility in the $T$ = 100-300 K range measured in $\mu_0 H$ = 500 G for the **NbMn₂** reference, plasma-treated, and exposed to 0% relative humidity samples.

| Sample name | First fitting stage | | Second fitting stage | | |
|---|---|---|---|---|---|
| | $J_{NbMn}$ [K] | $\chi_0 \times 10^{-3}$ [emu·mol⁻¹] | $J_{NbMn}$ [K] | $J_{MnMn}Z_{MnMn}$ [K] | $\chi_0 \times 10^{-3}$ [emu·mol⁻¹] |
| reference | -15.56(15) | -2.42(9) | -13.31(25) | 3.46(34) | -3.65(13) |
| plasma | -19.33(25) | -2.87(22) | -15.30(33) | 8.06(53) | -5.75(21) |
| 0% RH | -5.2(21.6) | -9.5(30) | -13.11(6) | -18.39(7) | -2.86(2) |

Both fitting stages for the reference and plasma-treated samples provided satisfactory results within the 100-300 K temperature range. However, for the post-DVS sample, only the second fitting stage could reproduce the $\chi^{-1}$ temperature dependence correctly. All fits predicted negative $J_{NbMn}$ values, indicating antiferromagnetic interaction between Nb$^{IV}$ and Mn$^{II}$ magnetic centers. In the first fitting stage, $J_{NbMn}$ was found to be -15.56(25) K for the reference sample, which then decreased by 24% to $J_{NbMn}$ = -19.33(25) K for the plasma-treated sample, or increased by 67% to $J_{NbMn}$ = -5.2(21.6) K for the post-DVS sample. However, the latter case's fitting error is large (over 400%). Therefore, the first fitting stage for the post-DVS sample is not reliable.

In the second fitting stage, the absolute values of $J_{NbMn}$ for the reference ($J_{NbMn}$ = -13.31(25) K) and plasma-treated ($J_{NbMn}$ = -15.30(33) K) samples were smaller by 15-20% compared to their corresponding values in the first stage. The ferromagnetic interactions between Mn$^{II}$ nearest neighbors as obtained from the fits: $Z_{MnMn}J_{MnMn}$ = 3.46(34) K for the reference and $Z_{MnMn}J_{MnMn}$ = 8.06(53) for the plasma-treated samples, effectively compensated for this difference, resulting in slightly better results than in the first stage. For the post-DVS sample, the second fitting stage yielded two comparable and competing negative exchange couplings, $Z_{NbMn}$ = -13.11(6) K and $Z_{MnMn}J_{MnMn}$ = -18.39(7) K. Such a situation would lead to magnetic frustration, and consequently, hindered the magnetization values as seen in Fig. 10. It should be noted that the applied model might not be suitable for the post-DVS sample due to potential crystallographic structure modification. Therefore, the obtained MFA model prediction for this sample should be taken carefully.

## 4. Discussion

The XRPD and magnetometry studies indicated that the core structure of the **NbMn₂** system and the isothermal magnetization process at $T$ = 2.0 K, including saturation magnetization, were preserved after plasma treatment. However, the interplane distances were slightly shortened due to this process. This led to an increase in the absolute value of exchange coupling $J_{NbMn}$ by 25% (in the first fitting stage) or an enhancement of antiferromagnetic $J_{NbMn}$ by 15% and ferromagnetic $Z_{MnMn}J_{MnMn}$ by 130% (in the second fitting stage). The enhancement shifted the transition temperature to the long-range magnetic order by almost 50%, from $T_C$ = 49 K to $T_C$ = 72 K.

Although no similar reports in the literature regarding molecular magnetism can be found, analogous observations have been made in nanoparticles or superconductors exposed to plasma. Plasma treatment in $Ni_{0.25}Zn_{0.75}Fe_2O_4$ nanoparticles increased the blocking temperature from 70 to 80 K and the superparamagnetic transition temperature from 240 to 250 K [27]. Similarly, in $Fe_{2.95}Zn_{0.05}O_4$ nanoparticles, plasma irradiation enhanced self-heating temperature from 65 to 70 °C for magnetic hyperthermia and blocking temperature from 95 to 105 K [26]. In both cases, the temperature shift was attributed to an increase in the internal magnetic energy of the material after exposure to plasma. Furthermore, an increase in critical temperature from 92 to 96 K was observed in $YBa_2Cu_3O_y$ [38] and from 83 up to 97 K in $Bi_2Sr_2CaCu_2O_{8+y}$ [39] ceramics after plasma treatment. The critical temperature enhancement was related to the absorption of active oxygen.

The changes in the magnetic properties of **NbMn₂** induced by plasma must have a different origin. Firstly, a new magnetic phase was created in plasma created from air, pure oxygen, nitrogen, and argon (Fig. 6b). Therefore, it is unlikely that specific atoms that were potentially absorbed onto the surface of the **NbMn₂** sample modified the magnetic properties. Secondly, two distinct magnetic phases with $T_C$ = 49 K and $T_C$ = 72 K were observed with no continuous transition, indicating there was no gradual accumulation of internal magnetic energy (Fig. 6a). Additionally, the reduction in crystallite size by 40% from $D$ = 98(16) nm for the reference sample to $D$ = 57(13) nm for the plasma-treated sample (as stated in the X-ray powder diffraction (XRPD) section) cannot be the dominant cause of the shift in $T_C$, as it would show continuous changes in $T_C$ for various plasma parameters (irradiation power and time). For example, the continuous increase of the crystallinity and magnetic moment with higher plasma power was found in iron oxide nanoparticles [19].

In our report, the new magnetic phase of **NbMn₂** with $T_C$ = 72 K was obtained exclusively by plasma treatment since the same phase could not be achieved by heating (90 and 150 °C), using ultraviolet light (Fig. 8) or exposing to 0% relative humidity in DVS (Fig. 11). In particular, the latter resulted in the $T_C$ = 40 K magnetic phase with a magnetic susceptibility temperature dependence that could not be reproduced by the MFA model (Fig. 11) assuming superexchange interactions present only between nearest neighboring $Mn^{II}$ and $Nb^{IV}$ centers ($Z_{MnMn}J_{MnMn} = 0$). In the DVS experiment (Fig. 4), all water of crystallization (four $H_2O$ molecules) and half of coordinated water (two $H_2O$ molecules) is probably leaving the **NbMn₂** system, causing the crystal lattice reorganization and thus alteration of its magnetic properties. Furthermore, the isothermal magnetization curve at $T = 2.0$ K (Fig. 10) for the post-DVS sample differs from the $M(\mu_0H)$ for reference and plasma-treated samples. The post-DVS sample has a coercivity of $\mu_0H_c = 300$ G and does not saturate up to $\mu_0H = 7$ T. In comparison, reference and plasma-treated samples have $\mu_0H_c = 20$-$60$ G, and their magnetization saturates at approximately $\mu_0H = 1$ T. The simple experiment of mixing **NbMn₂** powder with 1% $H_2O_2$ water solution resulted in creating the magnetic phase with $T_C = 40$ K and $\mu_0H_c = 300$ G, which are the same as for the post-DVS sample. The $H_2O_2$ solution may induce the desorption of the water molecules, suggesting that the observed magnetic properties in the post-DVS sample result from the removal of water molecules.

The increase in $T_C$ from 49 K to 72 K in the plasma-treated samples was probably due to the desorption of four or fewer molecules of water of crystallization while all four coordinated water molecules were maintained. This change in the crystal lattice structure may not be noticeable in the XRPD patterns (Fig. 2). Still, it could have caused the observed shrinkage of the crystallographic unit cell, leading to the modification of superexchange coupling constants. Moreover, the **NbMn₂** sample, which was grounded but not protected from air, showed a Curie temperature of $T_C = 64$ K (Fig. 5). This suggests that it is an intermediate magnetic phase between the reference and plasma-treated systems, where the improved surface-to-volume ratio caused partial removal of the water of crystallization.

The desorption experiments in other $Mn^{II}$ and $Nb^{IV}$-based octacyanidometalates support the above reasoning. In $\{[Mn^{II}(imH)]_2(H_2O)_4[Nb^{IV}(CN)_8]\cdot 4H_2O\}_n$, where imH = imidazole, Curie temperature shifted from $T_C = 25$ K to $T_C = 62$ K, due to the removal of all water molecules that destabilizes the hydrogen bonds system and, as a consequence, an additional cyano-bridge between $Mn^{II}$ and $Nb^{IV}$ was established [34,36]. Two-step dehydration was reported in $\{[Mn^{II}(pydz)(H_2O)_2][Mn^{II}(H_2O)_2][Nb^{IV}(CN)_8]\cdot 2H_2O\}_n$, where pydz = pyridazine, with $T_C = 43$, 68, and 98 K for as-synthesized, dehydrated (the loss of four $H_2O$ molecules), and anhydrous (the loss of two remaining water molecules) samples [32,37]. The first dehydration step causes hydrogen bonds destabilization and new cyano-bridge formation, while the second step involves alteration of the Mn ions coordination sphere, resulting in the shortening of selected bonds. The whole dehydration process involved unit cell contraction by up to 16%. Such conclusions correspond to the observations made in the plasma-treated **NbMn₂** samples.

## 5. Conclusions

Plasma irradiation and dehydration were used to create multiple magnetic phases from the three-dimensional **NbMn$_2$** molecular magnet. The study demonstrated that exposing the compound to air, oxygen, nitrogen, and argon-based plasma for a sufficient amount of time and with enough plasma-generating device power causes the transition temperature to the long-range magnetic order state shift from $T_C$ = 49 K to $T_C$ = 72 K. Plasma most probably induces desorption of the water of crystallization, shortening the Mn–Nb distances and enhancing the superexchange coupling constants by approximately 20%. As a result, the core crystallographic structure of the system is maintained, and only the Curie temperature is shifted. When the **NbMn$_2$** sample was exposed to atmosphere inside a measuring device, partial dehydration of the water of crystallization occurred, creating an intermediate state between the as-synthesized and plasma-treated samples, with $T_C$ = 64 K. Further desorption process removes water molecules attached to Mn$^{II}$ magnetic centers, causing major structure reorganization that decreased the Curie temperature to $T_C$ = 40 K. These findings demonstrate that plasma-induced modifications in molecular systems are a reliable and effective way to alter specific properties and obtain unique magnetic phases that would otherwise be unattainable.

## 6. References

1. Tyagi, P., Riso, C., Amir, U., Rojas-Dotti, C. & Martínez-Lillo, J. Exploring room-temperature transport of single-molecule magnet-based molecular spintronics devices using the magnetic tunnel junction as a device platform. *RSC Adv.* **10**, 13006–13015 (2020).

2. Fursina, A. A. & Sinitskii, A. Toward Molecular Spin Qubit Devices: Integration of Magnetic Molecules into Solid-State Devices. *ACS Appl. Electron. Mater.* **5**, 3531–3545 (2023).

3. Konieczny, P. *et al.* Magnetic cooling: a molecular perspective. *Dalt. Trans.* **51**, 12762–12780 (2022).

4. Kowalewska, P. & Szałowski, K. Magnetocaloric properties of V6 molecular magnet. *J. Magn. Magn. Mater.* **496**, 165933 (2020).

5. Günther, K., Grabicki, N., Battistella, B., Grubert, L. & Dumele, O. An All-Organic Photochemical Magnetic Switch with Bistable Spin States. *J. Am. Chem. Soc.* **144**, 8707–8716 (2022).

6. Huang, Y. *et al.* Pressure-controlled magnetism in 2D molecular layers. *Nat. Commun.* **14**, 3186 (2023).

7. Arczyński, M., Stanek, J., Sieklucka, B., Dunbar, K. R. & Pinkowicz, D. Site-Selective Photoswitching of Two Distinct Magnetic Chromophores in a Propeller-Like Molecule To Achieve Four Different Magnetic States. *J. Am. Chem. Soc.* **141**, 19067–19077 (2019).

8. Magott, M., Reczyński, M., Gaweł, B., Sieklucka, B. & Pinkowicz, D. A Photomagnetic Sponge: High-Temperature Light-Induced Ferrimagnet Controlled by Water Sorption. *J. Am. Chem. Soc.* **140**, 15876–15882 (2018).

9. Heczko, M., Reczyński, M., Näther, C. & Nowicka, B. Tuning of magnetic properties of the 2D CN-bridged $Ni^{II}$–$Nb^{IV}$ framework by incorporation of guest cations of alkali and alkaline earth metals. *Dalt. Trans.* **50**, 7537–7544 (2021).

10. Pacanowska, A., Reczyński, M. & Nowicka, B. Modification of Structure and Magnetic Properties in Coordination Assemblies Based on $[Cu(cyclam)]^{2+}$ and $[W(CN)8]^{3-}$. *Crystals* **9**, 45 (2019).

11. Pedersen, K. S., Bendix, J. & Clérac, R. Single-molecule magnet engineering: Building-block approaches. *Chem. Commun.* **50**, 4396–4415 (2014).

12. Czernia, D. *et al.* Influence of proton irradiation on the magnetic properties of two - dimensional Ni ( II ) molecular magnet. *Sci. Rep.* **13**, 14032 (2023).

13. Krupinski, M. *et al.* Magnetic transition from dot to antidot regime in large area Co/Pd nanopatterned arrays with perpendicular magnetization. *Nanotechnology* **28**, 085302 (2017).

14. Krupinski, M., Zarzycki, A., Zabila, Y. & Marszałek, M. Weak Antilocalization Tailor-Made by System Topography in Large Scale Bismuth Antidot Arrays. *Materials* **13**, 3246 (2020).

15. Laurano, R. *et al.* Plasma Treatment of Polymer Powder as an Effective Tool to Functionalize Polymers: Case Study Application on an Amphiphilic Polyurethane. *Polymers* **11**, 2109 (2019).

16. Wang, L. *et al.* Giant Stability Enhancement of $CsPbX_3$ Nanocrystal Films by Plasma-Induced Ligand Polymerization. *ACS Appl. Mater. Interfaces* **11**, 35270–35276 (2019).

17. Khelifa, F., Ershov, S., Habibi, Y., Snyders, R. & Dubois, P. Free-Radical-Induced Grafting from Plasma Polymer Surfaces. *Chem. Rev.* **116**, 3975–4005 (2016).

18. Wang, B. *et al.* Enhanced room temperature ferromagnetism in $MoS_2$ by N plasma treatment. *AIP Adv.* **10**, (2020).

19. Asghari, S., Mohammadi, M. A., Julaei, R. & Taheri, R. A. Surface Modification of Superparamagnetic Iron Oxide Nanoparticles by Argon Plasma for Medical Applications. *J. Appl. Biotechnol. Reports* **9**, 563–568 (2022).

20. Munir, M. A. *et al.* Microwave plasma treatment of NiCuZn ferrite nanoparticles: a novel approach of improving opto-physical and magnetic properties. *Appl. Phys. A* **128**, 345 (2022).

21. Fang, Z. *et al.* Enhanced ferromagnetic properties of $N_2$ plasma-treated carbon nanotubes. *J. Mater. Sci.* **54**, 2307–2314 (2019).

22. Li, D., Xu, L. M., Li, S. W. & Zhou, X. Plasma Treatment Enhanced Magnetic Properties in Manganese Doped Titanium Nitride Thin Films. *Chinese J. Chem. Phys.* **30**, 457–460 (2017).

23. Dutta, T. *et al.* Nitrogen plasma treatment in two-step temperature deposited FePt bilayer media. *J. Magn. Magn. Mater.* **461**, 6–13 (2018).

24. Haye, E. *et al.* Tuning the Magnetism of Plasma-Synthesized Iron Nitride Nanoparticles: Application in Pervaporative Membranes. *ACS Appl. Nano Mater.* **2**, 2484–2493 (2019).

25. Wang, W.-H., Chang, P.-C., Jiang, P. & Lin, W.-C. Plasma-induced magnetic patterning of FePd thin films without and with exchange bias. *Appl. Surf. Sci.* **527**, 146831 (2020).

26. Choi, H., Kim, C. S. & Kim, S. B. Magnetic properties and hyperthermia of Zn-doped $Fe_3O_4$ nanoparticles with plasma treatment. *J. Korean Phys. Soc.* **72**, 243–248 (2018).

27. Kim, H. J. & Choi, H. Effect of plasma treatment on magnetic properties and heating efficiency of Ni-Zn nanoparticles. *J. Magn. Magn. Mater.* **484**, 14–20 (2019).

28. Herrera, J. M. *et al.* Three-dimensional bimetallic octacyanidometalates $[M^{IV}\{(\mu$-$CN)_4Mn^{II}(H_2O)_2\}2 \cdot 4H_2O]_n$ (M=Nb, Mo, W): Synthesis, single-crystal X-ray diffraction and magnetism. *Comptes Rendus Chim.* **11**, 1192–1199 (2008).

29. Pinkowicz, D. *et al.* Nature of magnetic interactions in 3D {[M$^{II}$(pyrazole)$_4$]$_2$[Nb$^{IV}$(CN)$_8$]·4H$_2$O}$_n$ (M = Mn, Fe, Co, Ni) molecular magnets. *Inorg. Chem.* **49**, 7565–7576 (2010).

30. Willemin, S. *et al.* Crystal Structures and Intercalation Reactions of Three-Dimensional Coordination Polymers [M(H$_2$O)$_2$]$_2$ [Mo(CN)$_8$]·4H$_2$O (M = Co, Mn). *Eur. J. Inorg. Chem.* **2003**, 1866–1872 (2003).

31. Fitta, M., Pełka, R., Sas, W., Pinkowicz, D. & Sieklucka, B. Dinuclear molecular magnets with unblocked magnetic connectivity: Magnetocaloric effect. *RSC Adv.* **8**, 14640–14645 (2018).

32. Fitta, M. *et al.* Magnetocaloric Effect in a Mn$_2$-Pyridazine-[Nb(CN)$_8$] Molecular Magnetic Sponge. *Eur. J. Inorg. Chem.* **2012**, 3830–3834 (2012).

33. Fitta, M. *et al.* Magnetocaloric effect in M–pyrazole–[Nb(CN)$_8$] (M = Ni, Mn) molecular compounds. *J. Phys. Condens. Matter* **24**, 506002 (2012).

34. Fitta, M. *et al.* Magnetocaloric effect and critical behavior in Mn$_2$-imidazole-[Nb(CN)$_8$] molecular magnetic sponge. *J. Magn. Magn. Mater.* **396**, 1–8 (2015).

35. Konieczny, P. *et al.* Magnetic percolation in CN-bridged ferrimagnetic coordination polymers. *Dalt. Trans.* **47**, 11438–11444 (2018).

36. Pinkowicz, D. *et al.* Magnetic Spongelike Behavior of 3D Ferrimagnetic {[Mn$^{II}$(imH)]$_2$[Nb$^{IV}$(CN)$_8$]}$_n$ with $T_c$ = 62 K. *Inorg. Chem.* **47**, 9745–9747 (2008).

37. Pinkowicz, D. *et al.* Double switching of a magnetic coordination framework through intraskeletal molecular rearrangement. *Angew. Chemie - Int. Ed.* **50**, 3973–3977 (2011).

38. Chen, W. M., Jiang, S. S., Guo, Y. C., Liu, H. K. & Dou, S. X. Increase in $T_c$ of YBa$_2$Cu$_3$O$_y$ by oxygen plasma treatment. *Phys. C Supercond.* **341–348**, 2451–2452 (2000).

39. Yang, Z.-Q. *et al.* Plasma treatment of high-temperature superconducting materials. *Phys. C Supercond.* **282–287**, 1613–1614 (1997).

# 7. Acknowledgments

This work was supported by the Polish Minister of Education and Science (grant No.: DI2017 006047).

# 8. Author contributions

D.C.: study design, plasma treatment, magnetic properties data collection, analysis, and interpretation, manuscript preparation; P.K.: data interpretation, manuscript preparation; M.P.: powder X-ray data collection and analysis; B.N.: dynamic vapor sorption experiment; D.P.: sample synthesis, structure details.

# 9. Data availability

The datasets used and/or analyzed during the current study are available from the corresponding author upon reasonable request.

# 10. Additional information

The authors declare no competing interests.